\documentclass[prl,twocolumn]{revtex4-1}
\pdfoutput=0
\usepackage{graphicx}
\usepackage{amsmath,amssymb,amsthm,amsbsy}
\usepackage{latexsym}
\usepackage{amsfonts}
\usepackage{amssymb}

\newcommand{\Ai}{{\rm Ai}}
\newcommand{\Bi}{{\rm Bi}}

\begin{document}

\title{Constraints on the Lepton Charge, Spin and Mass from Quasi-Local Energy}
\author{Bjoern S. Schmekel}
\affiliation{Department of Physics, College of Studies for Foreign Diploma Recipients at the University of Hamburg, 20355 Hamburg, Germany}
\email{bss28@cornell.edu}

\begin{abstract}
The masses of the elementary particles as well as their charges and spins belong to the fundamental physical constants. 
Presently, no fundamental theory describing them is available, so their values remain mysterious.
In this work we offer an approach based on the Brown-York quasi-local energy which includes the self-energy of an object. In order to compute
this energy we model the spacetime of the renormalized electron (and other leptons) by the Kerr-Newman metric.
Placing conditions on the associated energies at different radii we arrive at various constraints on the mass, charge and spin which are derived from the Planck scale.
The large gap between the TeV-scale and the Planck energy is due to the highly non-trivial behavior of the used quasi-local potential. 
\end{abstract}

\maketitle
\section{Introduction}
Even though a unified theory combining quantum field theory and general relativity is still unavailable ultimately our conception of elementary
particles will have to be compatible with general relativity. As part of such an effort the question of the spacetime structure of elementary particles 
will have to be resolved as well. 
It has been known since a landmark paper by Carter \cite{PhysRev.174.1559} that the Kerr-Newman metric possesses many interesting characteristics including
a gyromagnetic ratio of $g=2$ which matches the value predicted for a fermion described by the Dirac equation.
Indeed there appears to be a deep relation between the Kerr-Newman metric and the Dirac equation which has been thoroughly investigated
by Burinskii \cite{Burinskii:2005mm,Burinskii:2007ur,Burinskii:2004qf,Burinskii_2020} and others, cf. e.g. \cite{Arcos:2002ip}. Apart from such deep interconnects on a 
more "macroscopic level" it should be possible to describe a particle viewed from some distance 
as a tiny rotating charged sphere at least to some approximation. According to the no-hair conjecture any black hole solution of the Einstein-Maxwell
equations can be completely characterized by its mass, charge and angular momentum with the spacetime being given by the Kerr-Newman metric.
The latter argument is not completely valid, though, because the Kerr-Newman metric describes a real black hole with an event horizon only if
$m^2 > a^2 + Q^2$, i.e. when the mass squared is much larger than the sum of the squares of the charge and the rotation parameter $a=J/m$
where $J$ is the angular momentum. However, for the electron we have in geometrized units with $G=c=1$ used throughout the entire text  
$m=6.76 \cdot 10^{-56} {\rm cm}$, $a=1.93 \cdot 10^{-11} {\rm cm}$ and $Q=1.38 \cdot 10^{-34} {\rm cm}$, i.e. the rotation parameter alone exceeds the mass by a 
factor of $2.86 \cdot 10^{44}$. Still, the Kerr-Newman metric naturally extends into this parameter space.

It has been argued that an electron cannot be a black hole because its Schwarzschild radius would be many order of magnitude smaller than
its Compton length thus not fitting inside its own Schwarzschild radius. There are two problems with this argument. First, it completely disregards
possible effects from the angular momentum and the charge which in our used unit system are much larger than the mass. Therefore, these
effects can almost certainly not be neglected. In fact, the Kerr-Newman metric used for the present purpose does not describe a black hole 
as mentioned before. Second, the notion of mass and energy is a complicated one in general relativity and one has
to settle on a suitable definition of energy before proceeding with arguments based on energy and mass. The notion is difficult because the
influence of gravity can always be gauged away at any single point in spacetime. Fortunately, a suitable definition is available. The Brown-York
quasilocal energy ("QLE")  \cite{Brown:1992br} satisfies all important properties an energy is supposed to possess. In particular it satisfies a conservation law
\cite{Schmekel:2018wbl} which attributes a change in QLE to a flux of ordinary stress-energy and a flux of a second quantity into or out of the region of interest. 
There is compelling evidence \cite{Schmekel:2018bcf,Oltean:2020mvt} that the latter is to be considered a flux of gravitational field energy. Furthermore, the QLE is additive in the sense
that the QLE of two disjoint regions is the sum of the QLEs contained in the regions. It is defined quasilocally, i.e. in a finite region larger than
a point thus avoiding the problem mentioned above and can be derived naturally from an action principle where all applicable boundary terms
\cite{PhysRevLett.28.1082} have been added to the Einstein-Hilbert action. As such the concept is fully covariant even though the exact value will depend on a chosen
time slicing and the size and shape of the boundary since the region cannot be point-like. Like potential energy in classical mechanics the QLE
allows to set a reference energy adding an arbitrary functional $S_0$ to the action which only depends on the induced metric of the chosen boundary.
Other than that the functional is arbitrary and has no influence on the conservation law and the equations of motion derived from the action.
With an appropriate reference term it is possible to recover the ADM limit, though. The reference term has been subject to a thorough
investigation in the literature because the suggested procedure to determine it proposed in the original article \cite{Brown:1992br} is not always applicable.  
The development of modifications to the original Brown-York QLE \cite{Liu:2003bx,Wang:2008jy} with a well-defined ADM limit can be seen as efforts to rectify this situation.
However, because of the reasons mentioned above the reference term is still arbitrary. As in past works we set $S_0=0$ which is a perfectly valid choice.
In any case we would expect a reasonable reference energy to be in the order of $r$ which can be neglected in our application.
We will either consider energy differences or argue that a suitable reference term will drop out in the small sphere limit. 
It is worth pointing out that the QLE is implemented as a surface integral over a contraction of a quasi-local stress-energy-momentum surface density.
Although evaluating the QLE gives the total energy consisting of ordinary stress-energy and gravitational field energy contained in a region enclosed by a boundary
it only requires knowledge of the metric and its derivatives on the boundary. Therefore, results for the QLE of the Kerr-Newman metric
will be valid as long as the Kerr-Newman metric as an exterior solution to the Einstein field equations is valid on the chosen boundary. 
For the interpretation of the following results it is important to stress that we ultimately apply the Kerr-Newman metric to the renormalized particle using
the renormalized values for charge and mass. We view the renormalized particle as surrounded by a cloud of virtual particles which are enclosed by
the boundary as well and which preserve the axial symmetry of the problem at least if the boundary exceeds a minimum size of the Compton length $r_c = h/m$. 
Like in quantum field theory the effect of the vacuum polarization is absorbed by replacing the bare charge and the bare mass with their renormalized counterparts.  
\section{QLE of the Kerr-Newman metric}
The QLE of the Kerr-Newman metric given in Boyer-Lindquist coordinates
\begin{eqnarray}  
ds^2=- \left ( 1-\frac{2mr-Q^2}{r^2+a^2 \cos^2 \theta} \right ) dt^2 + \nonumber \\
\frac{r^2+a^2 \cos^2 \theta}{r^2-2mr+a^2+Q^2} dr^2 +
\left (r^2 + a^2 \cos^2 \theta \right ) d \theta^2 + \nonumber \\
\sin^2 \theta \left (   r^2 + a^2 + \frac{\left ( 2mr -Q^2 \right ) a^2 \sin^2 \theta}{r^2 + a^2 \cos ^2 \theta}  \right ) d \phi^2 - \nonumber \\
\frac{2a \left ( 2mr -Q^2 \right ) \sin^2 \theta }{r^2 + a^2 \cos^2 \theta} d \phi dt  
\label{KNmetric}
\end{eqnarray}
has been computed before \cite{Schmekel:2018bcf}. The result is given by
\begin{eqnarray}
-i\frac{{6m{r^2} - 2{r^3} - r\left( {4{m^2} + 2{Q^2} + {a^2}} \right) + 2m{Q^2} + m{a^2}}}{{2\left| a \right|\sqrt {{Q^2} - 2mr + {r^2}} }}{\tilde \Xi _E} + \nonumber \\
i\frac{{m{r^2} + {r^3} - r\left( {4{m^2} - {a^2}} \right) + 2m{Q^2} + m{a^2}}}{{2\left| a \right|\sqrt {{Q^2} - 2mr + {r^2}} }}{\tilde \Xi _F} - \nonumber \\
\frac{{\left( {m - r} \right)\sqrt {\left( {{r^2} + {a^2}} \right)\left( {{Q^2} + {a^2} - 2mr + {r^2}} \right)} }}{{2{Q^2} - 4mr + 2{r^2}}} = E_1 \nonumber \\
 \label{Enoref}
\end{eqnarray}
with the time slice chosen by the unit normal vector $u^\mu = [(r^2+a^2\cos^2 \theta)/(r^2 -2mr +a^2 \cos^2 \theta + Q^2)]^{1/2} \delta_t^\mu $
%\begin{eqnarray}
%u^\mu = \sqrt{\frac{r^2+a^2\cos^2 \theta}{r^2 -2mr +a^2 \cos^2 \theta + Q^2}} \delta_t^\mu 
%\end{eqnarray}
and the boundary described by the unit normal vector $n^\mu = [(r^2+a^2 -2mr+Q^2)/(r^2+a^2 \cos^2 \theta)]^{1/2} \delta_r^\mu$
%\begin{eqnarray}
%n^\mu = \sqrt{\frac{r^2+a^2 -2mr+Q^2}{r^2+a^2 \cos^2 \theta}} \delta_r^\mu
%\end{eqnarray}
%Here, $\tilde \Xi_E \equiv \mathfrak{E} \left( i  | a/r  | , |r| (Q^2-2mr+r^2)^{-1/2} \right )$, $\tilde \Xi_F \equiv \mathfrak{F} \left( i  | a/r  | , |r| (Q^2-2mr+r^2)^{-1/2} \right )$

Here,
\begin{eqnarray}
\tilde \Xi_E \equiv \mathfrak{E} \left( i \left | \frac{a}{r} \right | , \frac{|r|}{\sqrt{Q^2-2mr+r^2}} \right ) \\
\tilde \Xi_F \equiv \mathfrak{F} \left( i \left | \frac{a}{r} \right | , \frac{|r|}{\sqrt{Q^2-2mr+r^2}} \right )
\end{eqnarray}
%and the incomplete elliptic integrals $\mathfrak{E}$ and $\mathfrak{F}$ are defined as in \cite{Schmekel:2018bcf}. 
and the incomplete elliptic integrals are defined as
\begin{eqnarray}
\mathfrak{E}(z,k) \equiv \int_0^z \frac{\sqrt{1-k^2 \zeta^2}}{\sqrt{1-\zeta^2}} d \zeta \\
\mathfrak{F}(z,k) \equiv \int_0^z \frac{1}{\sqrt{1-\zeta^2}\sqrt{1-k^2 \zeta^2}} d \zeta
\end{eqnarray}

Note that the spacetime possesses a ringlike singularity with radius $r_a=a$ in the associated Minkowski space.
%%% 
A subtlety arises when coupling the electromagnetic action (or any other action for a gauge field for that matter) to
the gravitational action as matter action. Loosley speaking the ordinary stress energy tensor can be derived by
varying the matter action with respect to the full spacetime metric whereas the quasilocal stress energy surface density 
is obtained by varying the full action with respect to the induced metric of the boundary. It is pointed out again
that the quasilocal energy contains both the energy due to the gravitational field and the contribution from ordinary
stress energy. In the original treatment \cite{Brown:1992br} it was assumed that the matter action depends on the full spacetime metric
only and not on the induced metric. However, once the matter action contains derivatives there may be a dependence on
the induced metric because the volume integral may be converted into a surface integral such that the variation of the
matter action with respect to the induced metric does not vanish as was pointed out in \cite{Booth:1999bn,Booth_1999,Booth:2000iq,Mondal:2022vmn}. 
The additional surface term like the reference term leads to a shift in the quasilocal energy. 
For several reasons this problem can presently be neglected. First, it appears that in the absence of an event horizon
as is assumed in our case the gauge of the vector potential can be chosen such that this additional contribution to the
QLE is zero. Even if an event horizon exists it may be possible to find a coordinate system such that coordinate singularity
at the event horizon vanishes allowing for a suitable electromagnetic gauge. However, the QLE remains dependent on the chosen gauge. 
It may be worth considering defining the quasi-local energy surface density as $\delta (S - S_m) / \delta \gamma_{ij}$ as a QLE
defined this way would still satisfy a conservation law \cite{Schmekel:2018bcf}.  
However, we simply choose to absorb any contributions from surface terms into a second reference term $S_0^{\rm gauge} [\gamma_{ij},A_{\mu}]$. 
Since this boundary term also depends on the field $A_{\mu}$ we have to make sure that a variation with respect to $A_{\mu}$ still reproduces
the correct equations of motion, i.e. the Maxwell equations. Applying the product rule gives
\begin{eqnarray}
S_m = S_{\rm EM} & = & - \frac{1}{4} \int_M d^4 x \sqrt{-g} F_{\mu \nu} F^{\mu \nu}  \\ \nonumber
                               & = & - \frac{1}{2} \int_M d^4 x \sqrt{-g} \nabla_{\mu} \left ( A_{\nu}  F^{\mu \nu} \right ) \\ \nonumber
                               & + & \frac{1}{2} \int_M d^4 x \sqrt{-g} A_{\nu} \nabla_{\mu} F^{\mu \nu}                              
\end{eqnarray}
where $F_{\mu \nu}$ can be written in terms of partial or covariant derivatives due to the antisymmetry of the tensor.
\begin{eqnarray}
F_{\mu \nu} F^{\mu \nu} = \left ( \nabla_{\mu} A_{\nu} - \nabla_{\nu} A_{\mu}  \right )  \left ( \partial^{\mu} A^{\nu} - \partial^{\nu} A^{\mu}  \right ) 
\end{eqnarray}
The total derivative yields a surface term which in standard  treatments of electrodynamics is held fixed by the von Neumann boundary
condition $\left . v_{\mu} F^{\mu \nu} \right |_{\partial M} = 0$.
We choose to subtract the emerging boundary term instead which gives the correct equations of motion when varying with respect to $A_{\mu}$
and avoids a further contribution to the QLE when varying with respect to $\gamma_{ij}$. 
In the non-Abelian case an additional contribution to the surface energy from the gauge field appears \cite{Mondal:2022vmn}
\begin{eqnarray}
g \int_B d^2 x \sqrt{\sigma} n_i g^{ik} A_{a0} f^a_{bc} A^b_0 A^c_k
\end{eqnarray}
where $g$ is the coupling constant and the indices $i$ and $k$ run over the spacelike coordinates describing the hypersurface $\Sigma$. 
This terms vanishes if $f^a_{bc} = - f^b_{ac}$, i.e. if the Lie algebra is a direct sum of simple compact Lie algebras. Since this requirement
is satisfies for the standard model of particle physics no new obstacles are expected.

To summarize we neglect contributions from both the reference term $S_0$ as well as from all other possible sources of surface terms
which may lead to a shift in QLE since these contributions do not alter the classical equations of motion.  
The picture may change once a full quantization of gravity is anticipated especially in non-trivial topologies. 
The problem deserves further investigation to understand which Lie algebras are permissible in the context of quasi-local energy. 
%%%%
\section{Constraints on Charge and Spin}
\begin{widetext}
\begin{figure}
\scalebox{0.8}{\includegraphics{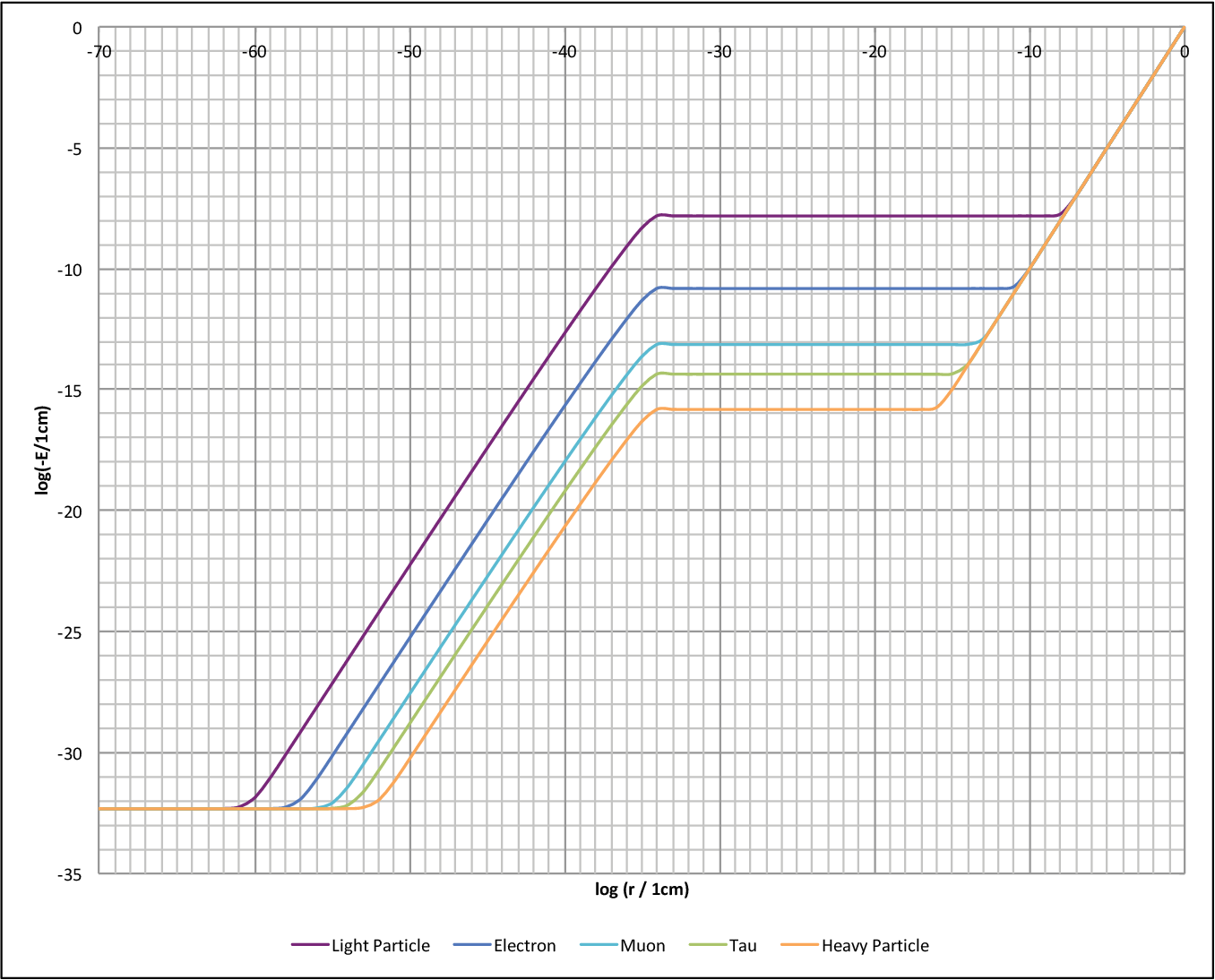}}
\caption{$\log (-E_1 / 1{\rm cm})$ vs. $\log (r / 1{\rm cm})$ for $J=\hbar/2=1.30 \cdot 10^{-66} {\rm cm}$, $| Q |= 1.38 \cdot 10^{-34} {\rm cm}$ and (from above to below)  $m_{\rm light} = 10^{-3} \cdot m_e$ ,
$m_e = 6.76 \cdot 10^{-56} {\rm cm}$,  $m_{\mu} = 206.77 \cdot m_e$, $m_{\tau} = 3477.23 \cdot m_e$ and finally $m_{\rm heavy} = 10^{5} \cdot m_e$. 
\footnote{A Mathematica notebook deriving and evaluating eqn. \ref{Enoref} can be found in \cite{SchmekelMathematica2020} }
}
\label{PlotQLE}
\end{figure}
\end{widetext}

Looking at the unreferenced QLE in fig. \ref{PlotQLE} we recognize a plateau extending roughly between $r_q < r < r_a$
where $r_q=Q=\sqrt{\alpha \hbar}$ and $\alpha$ is the fine structure constant. 
The value of $E_1$ at the plateau $E_{\rm plateau}$ is approximately $-r_a = -a = - J / m$. 
%
% Proof expanding the elliptic functions here
%
Demanding $r_a \approx r_c$ we obtain $J \sim h $ with the real value being $J = \hbar / 2 = h / (4 \pi)$. This is a
reasonable condition because we want the ring singularity to be hidden behind the Compton region.  
In the limit $r \longrightarrow0$ and $m \ll Q \ll a$ the unreferenced QLE $E_1$ can be approximated as
$E_1(r=0) \approx - \frac{m}{2} \left | a/Q \right | = - \frac{1}{2} \left | J/Q \right |$.
Substituting $a=J/m$ for the rotation parameter 
we obtain $E_1(r=0)=-2.926 \cdot E_{\rm Planck}$ which is independent of $m$ as long as the stated conditions are met. 
In  \cite{Schmekel:2018bcf} we argue that this result is not a coincidence. In fact it helps to resolve a puzzle because
the Compton wavelength of the Planck energy is identical to its associated Schwarzschild radius such that the Compton
wavelength can be confined within the Schwarzschild radius. Elevating this to a principle we have two conditions for
$J$ and $Q$ which allow us to solve for them. 

\section{Constraints on the Mass Spectrum}
The final remaining parameter we would like to determine is the (ADM) mass of the particle. Classically, we can only provide a large
range of allowed values. As mentioned above we treat leptons as overextreme Kerr-Newman "black holes" without event horizon
for which $m^2 < J^2 / m^2 + Q^2$ resulting in $m^2 < J$ for $J/m \gg Q$ and $m^2 < Q^2$ if $J=0$. Both conditions lead to approximately
the same bound of $m<10^{-34} {\rm cm}$ with angular momentum $J=\hbar/2$ or without. 
On the other hand it is obvious that in our model $m>0$ has to be demanded leaving us with a wide margin represented by the inequality $0<m<Q$. 
Otherwise, the size of the ring singularity $r_a = a = J/m$ would be infinite.   

Continuing with our analysis we consider a small ball with mass $M=E_1(r=0)$ trapped on the left side in the potential depicted in fig. \ref{PlotQLE}. 
The ball is trapped since both $M$ and the potential $E_1(r)$ are negative with $|M| < |E_1(r)|$. The uniformity of $E_1(r)$ below $r<r_s=2m$
leads us to regard $M$ as due to pure stress-energy located at $r=0$ classically because the Kerr-Newman metric is an electrovacuum solution
and gravitational field energy vanishes in the small sphere limit. 
(The picture is slightly more involved because for the Kerr-Newman metric $T_{\mu \nu} \neq 0$ even when $r>0$ due to the contribution to stress-energy
from the electromagnetic field. However, numerical integration gives $E_1^{\rm EM} (R) = 2 \pi \int_0^R dr \int_0^{\pi} d \theta \sqrt{h} \kappa^{-1} u^{\mu} u^{\nu} G_{\mu \nu} =
2 \pi \int_0^R dr \int_0^{\pi} d \theta \sqrt{h} u^{\mu} u^{\nu} T_{\mu \nu} \ll E_1(r=0)$ in the region $R < r_s$. )

Quantum mechanically this small pit of stress-energy at $r=0$ is now being spread out within the potential well. 
\subsection{Bound states}
\begin{figure}
\scalebox{0.6}{\includegraphics[trim={0 0 0 5cm},clip]{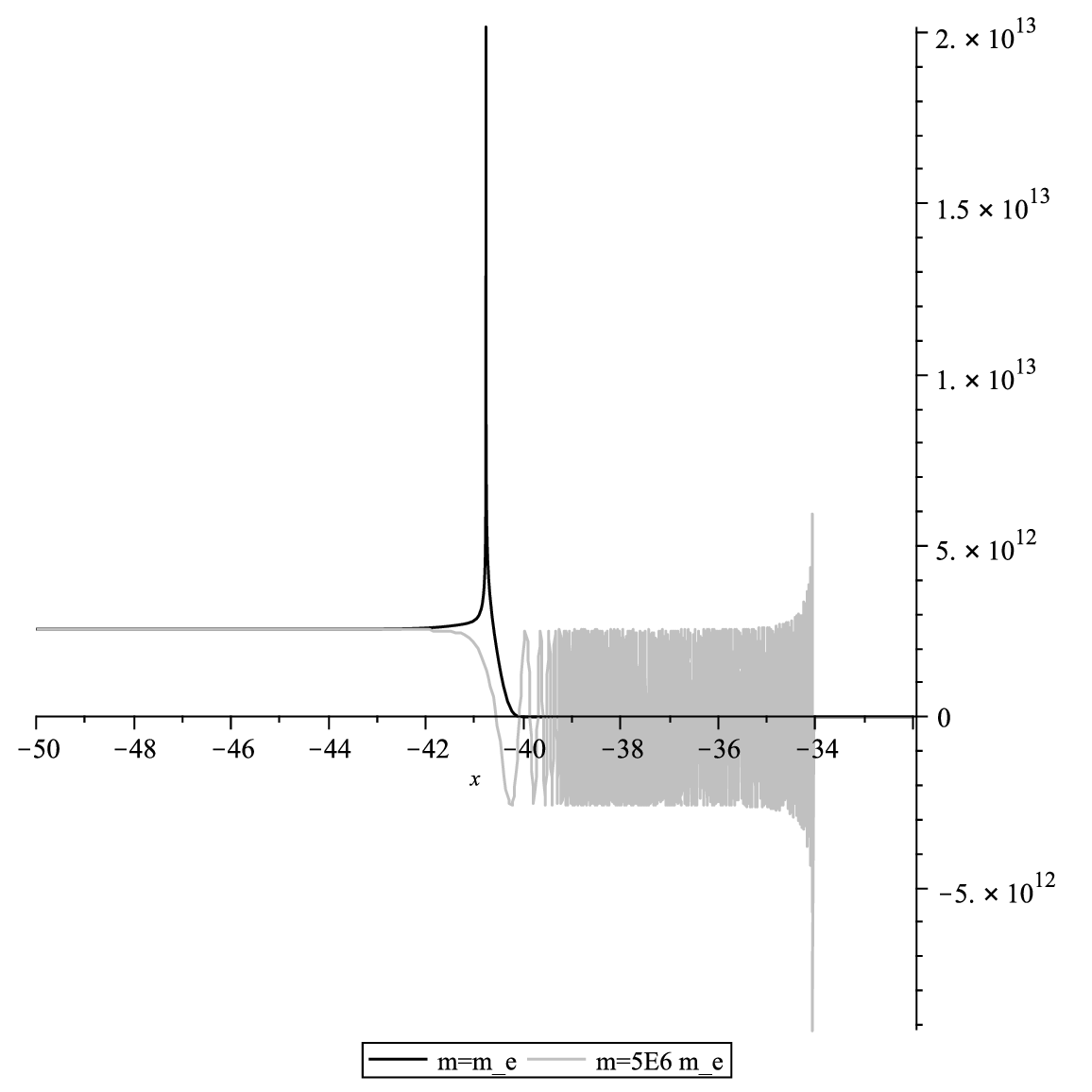}}
\caption{Real part of the unnormalized wavefunction $\Psi(r)=[2M(E-E_1(r))]^{-1/4} \cdot \exp \{i \hbar^{-1} \int^r dr' \sqrt{2M (E-E_1(r'))} \} $ in units of ${\rm cm}^{-1/2}$ vs. $x=\log(r/1{\rm cm})$ with the imaginary part being zero
as obtained from WKB approximation. }
\label{WKB}
\end{figure}

We approximate $E_1$ in the region $r_s<r<Q$ by $E_{1,{\rm slope}} (r) = - J r / (mQ) $ neglecting the plateau below $r_s$. 
Furthermore, we assume the dynamics of the system to be governed by a Schrodinger kind of equation
$ - \frac{\hbar^2}{2M} \frac{d^2}{dr^2} \Psi(r) + E_1(r) \Psi(r) = E \Psi(r) $
where all three quantities $M$, $E_1$ and $E$ are negative and the centrifugal term has been dropped. Needless to say this is a big assumption. In the absence of a theory of quantum gravity 
all subsequent calculations should be understood as order of magnitude estimates only. We note in passing that the metric in eqn. \ref{KNmetric} is approximately
in the geodesic gauge for $r \ll r_q$ in the direction $\theta=0$ which is a preferred gauge for the purpose of quantization \cite{Schmekel:2020iuv}. 
The solutions are given in terms of the Airy functions $\Ai$ and $\Bi$ discarding the latter such that suitable
boundary conditions can be applied. The energy eigenvalues of the bound states $E= \zeta_n E_0$ are associated with the energy scale $E_0= \hbar / (8Mm^2 \alpha) ^ {1/3}$
and the length scale $l_0=m^{1/3} M^{-1/3} \alpha^{1/6} \hbar^{1/2}$ where $\zeta_n$ are the roots of $\Ai$. 
For the three charged leptons we obtain
$l_{0,{\rm e}}=1.3 \cdot 10^{-41}  {\rm cm}$,
$l_{0,\mu}=7.7 \cdot 10^{-41}  {\rm cm}$ and
$l_{0,\tau}=2.0 \cdot 10^{-40}  {\rm cm}$
which are all very similar due to the presence of the cube root. Spotting these distances in fig. \ref{PlotQLE} they seem to approach the upper edge of the well, but
they are all well below $r_q$. The ratio of the potential energy at this location to the energy of the upper plateau is in the order of $10^{-6}$.
To maintain stability we do not want the ball to tunnel out of the potential well under any circumstances. 
In the WKB approximation shown in fig. \ref{WKB} the wave function drops to zero at $r=l_0$ for all practical purposes. 
Insisting on stability in all possible circumstances we also consider transitions from a lower mass state to a higher mass state. Assuming the potential $E_1$ changes according to
eqn. \ref{Enoref} or its approximations during a transition to a higher mass state with the energy eigenvalue remaining constant the wave function obtained by WKB approximation is
depicted in fig. \ref{WKB}. The wave function is getting uncomfortably close to the exit of the well at $r=r_q$. However, a mass ratio of roughly $1 : 5 \cdot 10^6$ could still be tolerated. 

Note that the condition $E<E_1(r_Q)$ only leads to the weak constraint $m<\alpha M$ where $\alpha M \sim Q$.
\subsection{Scattering states}
\begin{figure}
\scalebox{0.6}{\includegraphics{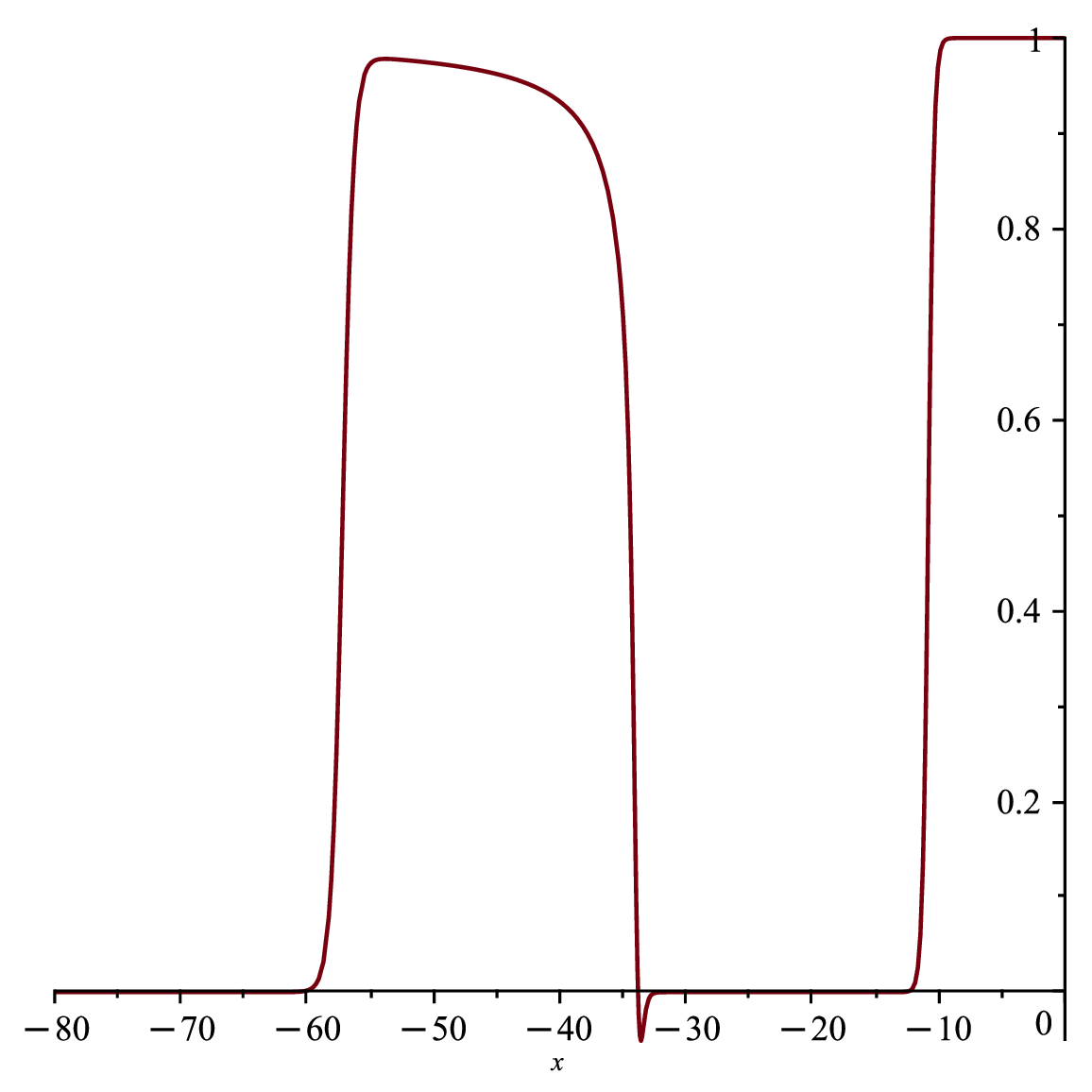}}
\caption{ $\frac{d}{dx} \log \left [ E_1 (1 {\rm cm} \cdot 10^x) \cdot 1 {\rm cm}^{-1} \right ]$ with $x=\log(r / 1 {\rm cm})$ } 
\label{PlotSlope}
\end{figure}

In this paragraph we use a slightly better approximation for the potential $E_1$ recognizing that the slope between the upper and lower plateau in the log-log diagramm in fig. \ref{PlotQLE} differs
slightly from $1$. 
Scattering states could possibly not only be contained within $r<r_q$ but within the much smaller range $r<r_s$ thus staying on the lower plateau such that crossing the barrier at $r=r_s$ is unlikely.
Classically, the fine structure constant is the ratio of the electrostatic force to the gravitational force for two particles with Planck mass and elementary charge.  
To make the forces between two ''halves'' of the particle at $r=0$ and $r=r_s$, respectively, equally strong the electrostatic repulsive force needs to be boosted by a factor of $\alpha^{-1}$. 
In order to determine a suitable threshold of the QLE we compare the field energy due to the electromagnetic and the gravitational field at $R=r_s$.
We have $E_1^{\rm EM} (R) = \tilde \alpha^{-1} \left [  E_1(R)-E_1(0) - E_1^{\rm EM} (R) \right ] $
where the term in the bracket is the energy contribution due to the gravitational field which is roughly in the order of $M$. Also, numerically we find that $\tilde \alpha$ is in the order of $\alpha$. 
Demanding a threshold of $E_1^{\rm EM} (R) \approx E_1(R) = \alpha^{-1} M$ and using the approximation $E_1(r) = c \alpha^{-\chi /2} (r/\sqrt{\hbar})^{\chi} \cdot \hbar / (2m)$ with $\chi \approx 0.976$ and $c=10.0$ 
the stated condition gives $m=0.016 \cdot m_e=6.8 \cdot 10^{-25} \cdot E_{\rm Planck}$. This suppression of the Planck energy by 24 orders of magnitude is a 
consequence of the peculiar properties of the used potential eqn. \ref{Enoref} with $E_1(r_s)$ scaling as $(m/\sqrt{\hbar})^{\chi-1}$ leaving an exponent much smaller than $1$
with $(m/\sqrt{\hbar})^{1-\chi}$ being in the order of unity for the observed mass spectrum. In fig.  \ref{PlotSlope} the exponent of the power law has been extracted with $0.976$ being the maximum value in the region of interest.
If nothing else this short calculation may provide an idea of how the quasi local potential might be able to bridge the large hierarchy between the Planck scale and the TeV-scale. 
\section{Conclusions}
The model presented in this letter may be criticized on several grounds. First, the model depends on the Brown-York QLE for a proper definition of energy and the Kerr-Newman metric
as a model for the spacetime structure of the particle. However, due to  the reasons mentioned above we consider these assumptions to be reasonable.
Second, the model is not completely self-consistent once quantum mechanical effects are taken into account. The Kerr-Newman metric
is an electrovacuum solution with a singular matter-energy distribution at $r=0$ while quantum mechanical effects spread out the probability density more or less within the well. 
Third, the requirement of stability may be regarded as ''anthropic'' . Fourth, the final theory of quantum gravity is still unknown. Still, it is hoped that the presented model
is a decent approximation allowing for some basic order of magnitude estimates.
Modeling the spacetime of the charged leptons by the Kerr-Newman metric we came up with two simple conditions constraining their angular momentum
and charge to be $J \sim \hbar$ and $Q \sim \sqrt{\hbar}$, respectively, ignoring small factors.
Since $\hbar$ is the only scale in our model with $m$ having the unit of length in our chosen unit system constraining the lepton masses is harder. While the inequality $0<m<Q$ follows quite naturally
stronger constraints are harder to come by because of the large hierarchy between the TeV-scale and the Planck scale. However, the situation is far from hopeless and a resolution of
this problem is not without precedent. After all the tiny Lamb shift is produced by suppressing the underlying scale by the fifth power of the fine structure constant.
In our case it was shown that a suppression of the Planck scale down to the TeV-scale may be accomplished by the used potential function which is subject to a power law with an exponent which differs only slightly from $1$. 
This led to an an optimal mass of $m=0.016 \cdot m_e$ for a stable particle with $J=\hbar/2$ and $|Q|=\sqrt{\alpha \hbar}$. Together with the mass range of  $1 : 5 \cdot 10^6$ we obtain the upper mass limit
$m<5 \cdot 10^6 \cdot m_e$. 
\acknowledgments
The author would like to acknowledge insightful discussions with James W. York, Jr.  and Alexander Burinskii. 
Also, support by Wolfram Research having provided assistance with Mathematica and free maintenance thereof is acknowledged
which resulted in \cite{SchmekelMathematica2020}. 
\bibliographystyle{apsrev4-1}
\bibliography{bib}

\begin{thebibliography}{19}%
\makeatletter
\providecommand \@ifxundefined [1]{%
 \@ifx{#1\undefined}
}%
\providecommand \@ifnum [1]{%
 \ifnum #1\expandafter \@firstoftwo
 \else \expandafter \@secondoftwo
 \fi
}%
\providecommand \@ifx [1]{%
 \ifx #1\expandafter \@firstoftwo
 \else \expandafter \@secondoftwo
 \fi
}%
\providecommand \natexlab [1]{#1}%
\providecommand \enquote  [1]{``#1''}%
\providecommand \bibnamefont  [1]{#1}%
\providecommand \bibfnamefont [1]{#1}%
\providecommand \citenamefont [1]{#1}%
\providecommand \href@noop [0]{\@secondoftwo}%
\providecommand \href [0]{\begingroup \@sanitize@url \@href}%
\providecommand \@href[1]{\@@startlink{#1}\@@href}%
\providecommand \@@href[1]{\endgroup#1\@@endlink}%
\providecommand \@sanitize@url [0]{\catcode `\\12\catcode `\$12\catcode
  `\&12\catcode `\#12\catcode `\^12\catcode `\_12\catcode `\%12\relax}%
\providecommand \@@startlink[1]{}%
\providecommand \@@endlink[0]{}%
\providecommand \url  [0]{\begingroup\@sanitize@url \@url }%
\providecommand \@url [1]{\endgroup\@href {#1}{\urlprefix }}%
\providecommand \urlprefix  [0]{URL }%
\providecommand \Eprint [0]{\href }%
\providecommand \doibase [0]{http://dx.doi.org/}%
\providecommand \selectlanguage [0]{\@gobble}%
\providecommand \bibinfo  [0]{\@secondoftwo}%
\providecommand \bibfield  [0]{\@secondoftwo}%
\providecommand \translation [1]{[#1]}%
\providecommand \BibitemOpen [0]{}%
\providecommand \bibitemStop [0]{}%
\providecommand \bibitemNoStop [0]{.\EOS\space}%
\providecommand \EOS [0]{\spacefactor3000\relax}%
\providecommand \BibitemShut  [1]{\csname bibitem#1\endcsname}%
\let\auto@bib@innerbib\@empty
%</preamble>
\bibitem [{\citenamefont {Carter}(1968)}]{PhysRev.174.1559}%
  \BibitemOpen
  \bibfield  {author} {\bibinfo {author} {\bibfnamefont {B.}~\bibnamefont
  {Carter}},\ }\href {\doibase 10.1103/PhysRev.174.1559} {\bibfield  {journal}
  {\bibinfo  {journal} {Phys. Rev.}\ }\textbf {\bibinfo {volume} {174}},\
  \bibinfo {pages} {1559} (\bibinfo {year} {1968})}\BibitemShut {NoStop}%
\bibitem [{\citenamefont {Burinskii}(2008{\natexlab{a}})}]{Burinskii:2005mm}%
  \BibitemOpen
  \bibfield  {author} {\bibinfo {author} {\bibfnamefont {A.}~\bibnamefont
  {Burinskii}},\ }\href {\doibase 10.1134/S0202289308020011} {\bibfield
  {journal} {\bibinfo  {journal} {Grav. Cosmol.}\ }\textbf {\bibinfo {volume}
  {14}},\ \bibinfo {pages} {109} (\bibinfo {year} {2008}{\natexlab{a}})},\
  \Eprint {http://arxiv.org/abs/hep-th/0507109} {arXiv:hep-th/0507109 [hep-th]}
  \BibitemShut {NoStop}%
%%CITATION = HEP-TH/0507109;%%
\bibitem [{\citenamefont {Burinskii}(2008{\natexlab{b}})}]{Burinskii:2007ur}%
  \BibitemOpen
  \bibfield  {author} {\bibinfo {author} {\bibfnamefont {A.}~\bibnamefont
  {Burinskii}},\ }\href {\doibase 10.1088/1751-8113/41/16/164069} {\bibfield
  {journal} {\bibinfo  {journal} {J. Phys. A}\ }\textbf {\bibinfo {volume}
  {41}},\ \bibinfo {pages} {164069} (\bibinfo {year} {2008}{\natexlab{b}})},\
  \Eprint {http://arxiv.org/abs/0710.4249} {arXiv:0710.4249 [hep-th]}
  \BibitemShut {NoStop}%
\bibitem [{\citenamefont {Burinskii}(2004)}]{Burinskii:2004qf}%
  \BibitemOpen
  \bibfield  {author} {\bibinfo {author} {\bibfnamefont {A.}~\bibnamefont
  {Burinskii}},\ }\href {\doibase 10.1103/PhysRevD.70.086006} {\bibfield
  {journal} {\bibinfo  {journal} {Phys. Rev. D}\ }\textbf {\bibinfo {volume}
  {70}},\ \bibinfo {pages} {086006} (\bibinfo {year} {2004})},\ \Eprint
  {http://arxiv.org/abs/hep-th/0406063} {arXiv:hep-th/0406063} \BibitemShut
  {NoStop}%
\bibitem [{\citenamefont {Burinskii}(2020)}]{Burinskii_2020}%
  \BibitemOpen
  \bibfield  {author} {\bibinfo {author} {\bibfnamefont {A.}~\bibnamefont
  {Burinskii}},\ }\href {\doibase 10.1088/1742-6596/1435/1/012053} {\bibfield
  {journal} {\bibinfo  {journal} {Journal of Physics: Conference Series}\
  }\textbf {\bibinfo {volume} {1435}},\ \bibinfo {pages} {012053} (\bibinfo
  {year} {2020})}\BibitemShut {NoStop}%
\bibitem [{\citenamefont {Arcos}\ and\ \citenamefont
  {Pereira}(2004)}]{Arcos:2002ip}%
  \BibitemOpen
  \bibfield  {author} {\bibinfo {author} {\bibfnamefont {H.}~\bibnamefont
  {Arcos}}\ and\ \bibinfo {author} {\bibfnamefont {J.}~\bibnamefont
  {Pereira}},\ }\href {\doibase 10.1023/B:GERG.0000046832.71368.a5} {\bibfield
  {journal} {\bibinfo  {journal} {Gen. Rel. Grav.}\ }\textbf {\bibinfo {volume}
  {36}},\ \bibinfo {pages} {2441} (\bibinfo {year} {2004})},\ \Eprint
  {http://arxiv.org/abs/hep-th/0210103} {arXiv:hep-th/0210103} \BibitemShut
  {NoStop}%
\bibitem [{\citenamefont {Brown}\ and\ \citenamefont
  {York}(1993)}]{Brown:1992br}%
  \BibitemOpen
  \bibfield  {author} {\bibinfo {author} {\bibfnamefont {J.}~\bibnamefont
  {Brown}}\ and\ \bibinfo {author} {\bibfnamefont {J.}~\bibnamefont {York},
  \bibfnamefont {James~W.}},\ }\href {\doibase 10.1103/PhysRevD.47.1407}
  {\bibfield  {journal} {\bibinfo  {journal} {Phys. Rev. D}\ }\textbf {\bibinfo
  {volume} {47}},\ \bibinfo {pages} {1407} (\bibinfo {year} {1993})},\ \Eprint
  {http://arxiv.org/abs/gr-qc/9209012} {arXiv:gr-qc/9209012} \BibitemShut
  {NoStop}%
\bibitem [{\citenamefont {Schmekel}(2018)}]{Schmekel:2018wbl}%
  \BibitemOpen
  \bibfield  {author} {\bibinfo {author} {\bibfnamefont {B.~S.}\ \bibnamefont
  {Schmekel}},\ }\href {\doibase 10.1103/PhysRevD.98.104051} {\bibfield
  {journal} {\bibinfo  {journal} {Phys. Rev. D}\ }\textbf {\bibinfo {volume}
  {98}},\ \bibinfo {pages} {104051} (\bibinfo {year} {2018})},\ \Eprint
  {http://arxiv.org/abs/1807.07550} {arXiv:1807.07550 [gr-qc]} \BibitemShut
  {NoStop}%
\bibitem [{\citenamefont {Schmekel}(2019)}]{Schmekel:2018bcf}%
  \BibitemOpen
  \bibfield  {author} {\bibinfo {author} {\bibfnamefont {B.~S.}\ \bibnamefont
  {Schmekel}},\ }\href {\doibase 10.1103/PhysRevD.100.124011} {\bibfield
  {journal} {\bibinfo  {journal} {Phys. Rev. D}\ }\textbf {\bibinfo {volume}
  {100}},\ \bibinfo {pages} {124011} (\bibinfo {year} {2019})},\ \Eprint
  {http://arxiv.org/abs/1811.03551} {arXiv:1811.03551 [gr-qc]} \BibitemShut
  {NoStop}%
\bibitem [{\citenamefont {Oltean}\ \emph {et~al.}(2021)\citenamefont {Oltean},
  \citenamefont {Bazrafshan~Moghaddam},\ and\ \citenamefont
  {Epp}}]{Oltean:2020mvt}%
  \BibitemOpen
  \bibfield  {author} {\bibinfo {author} {\bibfnamefont {M.}~\bibnamefont
  {Oltean}}, \bibinfo {author} {\bibfnamefont {H.}~\bibnamefont
  {Bazrafshan~Moghaddam}}, \ and\ \bibinfo {author} {\bibfnamefont {R.~J.}\
  \bibnamefont {Epp}},\ }\href {\doibase 10.1088/1361-6382/abeae3} {\bibfield
  {journal} {\bibinfo  {journal} {Class. Quant. Grav.}\ }\textbf {\bibinfo
  {volume} {38}},\ \bibinfo {pages} {085012} (\bibinfo {year} {2021})},\
  \Eprint {http://arxiv.org/abs/2006.10068} {arXiv:2006.10068 [gr-qc]}
  \BibitemShut {NoStop}%
\bibitem [{\citenamefont {York}(1972)}]{PhysRevLett.28.1082}%
  \BibitemOpen
  \bibfield  {author} {\bibinfo {author} {\bibfnamefont {J.~W.}\ \bibnamefont
  {York}, \bibfnamefont {Jr.}},\ }\href {\doibase 10.1103/PhysRevLett.28.1082}
  {\bibfield  {journal} {\bibinfo  {journal} {Phys. Rev. Lett.}\ }\textbf
  {\bibinfo {volume} {28}},\ \bibinfo {pages} {1082} (\bibinfo {year}
  {1972})}\BibitemShut {NoStop}%
\bibitem [{\citenamefont {Liu}\ and\ \citenamefont {Yau}(2003)}]{Liu:2003bx}%
  \BibitemOpen
  \bibfield  {author} {\bibinfo {author} {\bibfnamefont {C.-C.~M.}\
  \bibnamefont {Liu}}\ and\ \bibinfo {author} {\bibfnamefont {S.-T.}\
  \bibnamefont {Yau}},\ }\href {\doibase 10.1103/PhysRevLett.90.231102}
  {\bibfield  {journal} {\bibinfo  {journal} {Phys. Rev. Lett.}\ }\textbf
  {\bibinfo {volume} {90}},\ \bibinfo {pages} {231102} (\bibinfo {year}
  {2003})},\ \Eprint {http://arxiv.org/abs/gr-qc/0303019} {arXiv:gr-qc/0303019}
  \BibitemShut {NoStop}%
\bibitem [{\citenamefont {Wang}\ and\ \citenamefont {Yau}(2009)}]{Wang:2008jy}%
  \BibitemOpen
  \bibfield  {author} {\bibinfo {author} {\bibfnamefont {M.-T.}\ \bibnamefont
  {Wang}}\ and\ \bibinfo {author} {\bibfnamefont {S.-T.}\ \bibnamefont {Yau}},\
  }\href {\doibase 10.1103/PhysRevLett.102.021101} {\bibfield  {journal}
  {\bibinfo  {journal} {Phys. Rev. Lett.}\ }\textbf {\bibinfo {volume} {102}},\
  \bibinfo {pages} {021101} (\bibinfo {year} {2009})},\ \Eprint
  {http://arxiv.org/abs/0804.1174} {arXiv:0804.1174 [gr-qc]} \BibitemShut
  {NoStop}%
\bibitem [{\citenamefont {Booth}\ and\ \citenamefont
  {Mann}(1999{\natexlab{a}})}]{Booth:1999bn}%
  \BibitemOpen
  \bibfield  {author} {\bibinfo {author} {\bibfnamefont {I.}~\bibnamefont
  {Booth}}\ and\ \bibinfo {author} {\bibfnamefont {R.~B.}\ \bibnamefont
  {Mann}},\ }\href {\doibase 10.1063/1.1301583} {\bibfield  {journal} {\bibinfo
   {journal} {AIP Conf. Proc.}\ }\textbf {\bibinfo {volume} {493}},\ \bibinfo
  {pages} {182} (\bibinfo {year} {1999}{\natexlab{a}})},\ \Eprint
  {http://arxiv.org/abs/gr-qc/9907077} {arXiv:gr-qc/9907077} \BibitemShut
  {NoStop}%
\bibitem [{\citenamefont {Booth}\ and\ \citenamefont
  {Mann}(1999{\natexlab{b}})}]{Booth_1999}%
  \BibitemOpen
  \bibfield  {author} {\bibinfo {author} {\bibfnamefont {I.~S.}\ \bibnamefont
  {Booth}}\ and\ \bibinfo {author} {\bibfnamefont {R.~B.}\ \bibnamefont
  {Mann}},\ }\href {\doibase 10.1103/physrevd.60.124009} {\bibfield  {journal}
  {\bibinfo  {journal} {Phys. Rev. D}\ }\textbf {\bibinfo {volume} {60}}
  (\bibinfo {year} {1999}{\natexlab{b}}),\
  10.1103/physrevd.60.124009}\BibitemShut {NoStop}%
\bibitem [{\citenamefont {Booth}(2000)}]{Booth:2000iq}%
  \BibitemOpen
  \bibfield  {author} {\bibinfo {author} {\bibfnamefont {I.~S.~N.}\
  \bibnamefont {Booth}},\ }\emph {\bibinfo {title} {{A Quasilocal Hamiltonian
  for gravity with classical and quantum applications}}},\ \href@noop {}
  {\bibinfo {type} {Other thesis}} (\bibinfo {year} {2000}),\ \Eprint
  {http://arxiv.org/abs/gr-qc/0008030} {arXiv:gr-qc/0008030} \BibitemShut
  {NoStop}%
\bibitem [{\citenamefont {Mondal}\ and\ \citenamefont
  {Yau}(2022)}]{Mondal:2022vmn}%
  \BibitemOpen
  \bibfield  {author} {\bibinfo {author} {\bibfnamefont {P.}~\bibnamefont
  {Mondal}}\ and\ \bibinfo {author} {\bibfnamefont {S.-T.}\ \bibnamefont
  {Yau}},\ }\href {\doibase 10.1103/PhysRevD.105.104068} {\bibfield  {journal}
  {\bibinfo  {journal} {Phys. Rev. D}\ }\textbf {\bibinfo {volume} {105}},\
  \bibinfo {pages} {104068} (\bibinfo {year} {2022})},\ \Eprint
  {http://arxiv.org/abs/2201.12956} {arXiv:2201.12956 [gr-qc]} \BibitemShut
  {NoStop}%
\bibitem [{\citenamefont {Schmekel}(2020{\natexlab{a}})}]{Schmekel:2020iuv}%
  \BibitemOpen
  \bibfield  {author} {\bibinfo {author} {\bibfnamefont {B.~S.}\ \bibnamefont
  {Schmekel}},\ }\href@noop {} {\  (\bibinfo {year} {2020}{\natexlab{a}})},\
  \Eprint {http://arxiv.org/abs/2009.00264} {arXiv:2009.00264 [gr-qc]}
  \BibitemShut {NoStop}%
\bibitem [{\citenamefont
  {Schmekel}(2020{\natexlab{b}})}]{SchmekelMathematica2020}%
  \BibitemOpen
  \bibfield  {author} {\bibinfo {author} {\bibfnamefont {B.~S.}\ \bibnamefont
  {Schmekel}},\ }\href@noop {} {\enquote {\bibinfo {title} {Quasi-local energy
  of a charged rotating object described by the kerr-newman metric},}\
  }\bibinfo {howpublished} {The Notebook Archive, Wolfram Foundation} (\bibinfo
  {year} {2020}{\natexlab{b}}),\ \bibinfo {note}
  {https://notebookarchive.org/2020-05-cglhk3s}\BibitemShut {NoStop}%
\end{thebibliography}%

%\appendix
%\section{Appendix A}

\end{document}